\documentclass[conference]{IEEEtran}
\ifCLASSINFOpdf
  \usepackage[pdftex]{graphicx}
  \DeclareGraphicsExtensions{.pdf,.jpeg,.png}
\fi
\usepackage{amssymb}
\usepackage[cmex10]{amsmath}
\usepackage{cite}
\usepackage{mathtools}
\usepackage{soul,color}
\usepackage{multirow}
\usepackage{algorithm}
\usepackage{algorithmic}
\usepackage{bm}
\usepackage{enumerate}
\usepackage{cite}

\usepackage{algorithmic}
\usepackage{algorithm}

\usepackage{array}
\ifCLASSINFOpdf   
   \usepackage[pdftex]{graphicx}     
   \graphicspath{../Simulation_results/}    
   \DeclareGraphicsExtensions{.pdf,.jpeg,.png,.jpg}   
 \else   
   \usepackage[dvips]{graphicx}   
   \graphicspath{{../eps/}}    
\fi
\ifCLASSINFOpdf
\else
   \usepackage[dvips]{graphicx}
   \DeclareGraphicsExtensions{.eps}
\fi
\sethlcolor{yellow}

\usepackage{array}
\newtheorem{theorem}{Theorem}[section]

\newtheorem{lemma}[theorem]{Lemma}

\DeclarePairedDelimiter\abs{\lvert}{\rvert}%
\DeclarePairedDelimiter\norm{\lVert}{\rVert}%

\makeatletter
\let\oldabs\abs
\def\abs{\@ifstar{\oldabs}{\oldabs*}}
\let\oldnorm\norm
\def\norm{\@ifstar{\oldnorm}{\oldnorm*}}
\makeatother

\DeclareMathOperator*{\argmin}{arg\,min}

\begin{document}
%
\title{Per-Tone model for Common Mode sensor based alien noise cancellation for Downstream xDSL}





\author{\IEEEauthorblockN{Ramanjit Ahuja\IEEEauthorrefmark{1},
Pravesh Biyani\IEEEauthorrefmark{2},
Surendra Prasad\IEEEauthorrefmark{1}\IEEEauthorrefmark{3},
Brejesh Lall\IEEEauthorrefmark{1}\IEEEauthorrefmark{3}}
\IEEEauthorblockA{\IEEEauthorrefmark{1}Department of Electrical Engineering, Indian Institute of Technology,
New Delhi, India}
\IEEEauthorblockA{\IEEEauthorrefmark{2}Indraprastha Institute of Information Technology, New Delhi, India}
\IEEEauthorblockA{\IEEEauthorrefmark{3}Bharti School of Telecom Technology and Management, Indian Institute of Technology,
New Delhi, India}

}


%


\maketitle

\begin{abstract}
For xDSL systems, alien noise cancellation using an additional common mode sensor at 
the downstream receiver can be thought of as interference cancellation in
a Single Input Dual Output (SIDO) system. The coupling between the common mode and differential mode can be modelled as an LTI
system with a long impulse response, resulting in high complexity for cancellation. Frequency domain per-tone cancellation offers a low complexity approach to the problem besides having other advantages like faster training, but suffers from loss in cancellation performance due to approximations in the per-tone model. We analyze this loss and show that it is possible to minimize it by a convenient post-training ``delay" adjustment. We also show via measurements that the loss of cancellation performance due to the per-tone model is not very large for real scenarios.
\end{abstract}
%
\IEEEpeerreviewmaketitle

\section{Introduction} \label{sec:Introduction}
Cancellation of alien noise in xDSL receivers has attracted significant interest recently \cite{Henkel_1}\cite{dsl_plc}\cite{cm_perforlimit}\cite{broadcom_patent}. Sources of electromagnetically coupled alien noises include PLC modems, appliances (e.g. treadmill) and switching power supplies \cite{AEEE942}\cite{dsl_plc}. Mitigating the impact of these noises by cancellation can be done using an additional sensor, e.g. a common mode (CM) sensor or an unused twisted pair\cite{Magesacher_1}\cite{rficancel_jsac}. In xDSL, while differential mode (DM) is used for transmitting the data signal over the unshielded twisted pair, electro-magnetically coupled noises appear as CM signals{}. Ideally the CM and DM transmission modes are isolated but in practice, due to cable imbalances, there exist significant leakages between the CM and DM which is how alien/impulse noise leaks into the differential mode signal. The leakage couplings between CM and DM can be modelled as linear time-invariant (LTI) systems \cite{rficancel_jsac} with very long impulse responses \cite{Henkel_1}\cite{icc}.

Alien noise cancellation can be effected by spatial whitening via optimum linear combination of the CM and DM signals. However these involve high complexity due to the long impulse response if the couplings are modelled as finite impulse response (FIR) systems. Moreover we need to estimate the coupling between CM and DM to derive the linear canceller and this coupling usually needs to be estimated in the presence of a strong useful data signal in DM. This is because estimating the cross-correlation requires that CM be excited by the alien/impulse noise events and noise events may start only after the modem has trained up.

The frequency domain per-tone model of cancellation alluded to in \cite{broadcom_patent}\cite{patent}\cite{dsl_plc}, has low-complexity, easily incorporates decision-directed estimation and hence faster training\cite{icc} and implementation convenience. In the per-tone model, it is assumed that the alien noise in CM  undergoes a circular convolution with the CM-DM coupling function, while in reality the convolution is linear since the alien noise does not have a cyclic structure. This approximation introduces a penalty on the cancellation performance and per-tone cancellation residual noise can be potentially inferior to the Cramer-Rao lower bound (CRLB) for the residual noise.

  \par 
Following are the main contributions of the paper:
 \begin{itemize}
 \item We provide analytical treatment of the impact of non-cyclic structure of the alien noise signal and quantify the expected loss of performance.
\item We outline a method to optimize this loss by a post-training adjustment and we also suggest a method to derive a near- optimal time-domain canceller using the per-tone canceller coefficients.
 \end{itemize}
 \subsubsection*{Notations}
Time domain signals,impulse response etc. are denoted by lower-case letters, e.g. $y(n)$. Frequency domain signals are denoted by upper-case letters e.g. $Y_d(q),Y_c$. Matrices are denoted by bold letters while vectors are denoted by bold italicized letters. 
Superscript $^{*}$ denotes a conjugate.

\section{System Model and Proposed Canceller}

\begin{figure}
\includegraphics[clip, trim=2.8cm 15cm 3cm 1.5cm, width=0.5\textwidth]{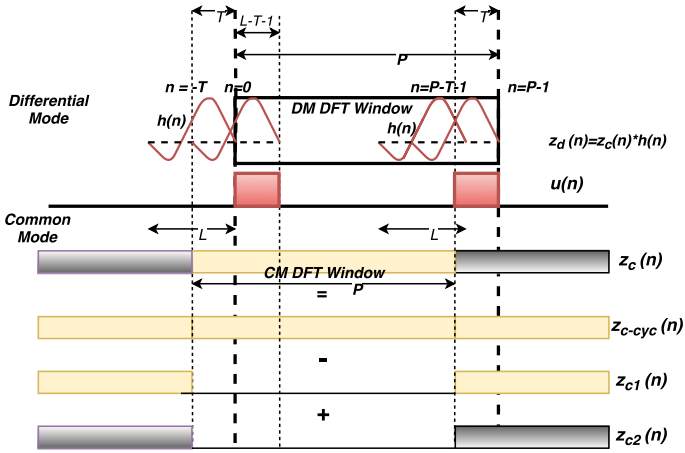}
\caption{CM-DM coupling illustrated showing the convolution of the coupling impulse response and the resulting uncancellable signal }
\label{fig:cm_dm_convolution}
\end{figure}
\subsection{System Model}
We assume that the CM-DM coupling modelled as an LTI time domain FIR filter \cite{Henkel_1}\cite{rficancel_jsac} denoted by $h(n)$ is $L$ taps long. Let $x(n)$ be the cyclically extended time domain DMT modulated data signal and $h_{d}(n)$ be the $M$ tap long impulse response of the DM direct channel between the CO transmitter and the CPE receiver. The time domain alien noise signals in the CM and DM are denoted by $z_{c}(n)$ and $z_{d}(n)$ while $v_{c}(n)$ and $v_{d}(n)$ constitute the background AWGN noise at the two sensors. There also exists a reverse DM to CM coupling which results in the DM data signal $x(n)$ leaking\footnote{This leakage signal is extremely small in comparison to the received DM useful data signal (appx 50 dB attenuation) \cite{Henkel_1} (also confirmed by lab measurements) and will not lead to any significant performance gain if combined with the DM data signal. The formulation can be modified to include this if needed but is skipped in interest of simplicity.} into the CM but is ignored in our formulation for simplicity.  The received time domain signals in CM and DM, $y_{c}(n)$ and $y_{d}(n)$ are given by:
\begin{flalign}
y_{c}(n) &= z_{c}(n) + v_{c}(n)\\
y_{d}(n) &= \sum_{k=0}^{k=M-1}{h_{d}(k)x(n-k)} + z_{d}(n) + v_{d}(n) \\
z_{d}(n) &= \sum_{k=0}^{k=L-1}{h_{cd}(k)z_{c}(n-k)}\label{eqn:conv}
\end{flalign}
A discrete Fourier transform (DFT) of length $P$ is applied to both the CM and DM real-valued 
time domain signal blocks ( VDSL transmission is baseband ) $y_{c}(n)$ and $y_{d}(n)$, resulting in corresponding complex valued hermitian-symmetric frequency domain signals. The corresponding frequency domain representations for the $q^{th}$ tone are given by :
\begin{flalign}\label{eqn:freq_dom_dm}
Y_{c}(q) &= Z_{c}(q)+ V_{c}(q)\\
Y_{d}(q) &= H_{d}(q)X(q) + Z_{d}(q) + V_{d}(q)
\
\end{flalign}
where,  $H_{d}(q)$ is the DM channel coefficient and $X(q)$ is the transmitted data symbol. We will now omit the tone index $q$ for rest of the paper wherever it is unambiguous to do so.
The alien noise component on any tone in DM, i.e. $Z_{d}$ can be split up as:
\begin{equation}
Z_{d}(q)=H_{pertone}(q)Z_{c}(q)+U(q)
\end{equation}
where $U$ and  $Z_{c}$ are uncorrelated and are both independent of $V_{d}$. The uncorrelated term $U$ consisting of contributions from neighbouring tones as well as from outside the signal window used for DFT of $z_{c}(n)$ represents the 
uncancellable (in the per-tone model) portion of the alien noise signal and arises because $z_{d}(n) = z_{c}(n)*h(n)$ and $z_{c}(n)$ is not cyclic.
We define the following terms for each tone:
\begin{gather*}
 I_{cm} = E\{\abs{Z_{c}}^2\}, \, I_{dm} = E\{\abs{Z_{d}}^2\},\, I_{U}=E\{\abs{U}^2\}\\
 \sigma_{X}^2=E\{\abs{X}^2\},\sigma_{V_{c}}^2=E\{\abs{V_{c}}^2\},\,\sigma_{V_{d}}^2=E\{\abs{V_{d}}^2\},\, \delta=\frac{\sigma_{V_{c}}^2}{I_{cm}}
\end{gather*}
\par

\subsection{Proposed Per-Tone Canceller}
The proposed per-tone canceller consists of linearly combining the received frequency domain CM and DM signals tone-wise with a different cancellation coefficient $\beta(q)$ used for each tone.
\begin{equation} \label{eqn:combiner}
\tilde Y_d=Y_d-\beta \,Y_c.
\end{equation}
The optimal cancellation coefficient obtained by minimising the residual error variance after cancellation is given by:
\begin{equation} \label{eqn:moe_form}
\beta_{opt} = \frac{E\{Y_dY_c^{*}\}}{E\{ \abs{Y_c}^2\}}=\frac{H_{pertone}}{1+\delta}
\end{equation}
The variance of the residual noise on any tone after cancellation by the optimum linear canceller designated the Per-Tone Lower Bound (PTLB) is given by: \begin{equation}
\Omega_{PTLB} =
I_{dm}\left(\frac{\delta}{1+\delta}\right)^{2}+I_{U}+\sigma_{V_d}^{2}+\frac{|H_{pertone}|^2}{(1+\delta)^2}\sigma_{V_c}^{2}
\end{equation}.  
The impact of uncancellable noise on the cancellation performance(vis a vis linear time domain cancellation) is small if the folded in CM background noise dominates the residual noise i.e. $\frac{|H_{pertone}|^2}{(1+\delta)^2}\sigma_{V_c}^{2} >> I_U$ which happens if the coupling $H_{cd}$ is large or if the CM background noise power is very high.
%

%
%
%
\section{Analysis of Per-Tone Cancellation Model}
We now derive an analytical expression for the coefficient $H_{pertone}(q)$, which should be equal to the DFT of $h_{cd}(n)$ in case $z_c(n)$ is cyclic. 
Figure \ref{fig:cm_dm_convolution} shows the CM and DM time domain signal blocks over which the DFT is performed and the two blocks are misaligned by $T$ samples. Note that the DM alien noise signal $z_{d}(n)$ which is modelled as a convolution of $z_c(n)$ and $h_{cd}(n)$ includes portions of $z_c(n)$ which lie outside the CM signal block used for the DFT due to the delay spread of $h_{cd}(n)$ and also due to the misalignment $T$.
\par
Consider a decomposition of $z_{c}(n)$ , $z_{c}(n)=z_{c-cyc}(n)+z_{c1}(n)+z_{c2}(n)$ as shown in figure \ref{fig:cm_dm_convolution} where $z_{c-cyc}(n)$ is a cyclically extended version of the CM DFT block i.e. $z_{c-cyc}(n) = z_{c}(n)$ for $-T \leq n \leq P-T-1$ while  $z_{c1}(n) = -z_{c-cyc}(n)$ for $n<-T$ and $n > P-T-1$ and $z_{c1}(n) = 0$ otherwise. Similarly $z_{c2}(n) = z_{c}(n)$ for $n<-T$ and $n > P-T-1$ and $z_{c2}(n) = 0$ otherwise.
The portion of $z_{c1}(n)$ and $z_{c2}(n)$ which folds in into the DFT block for $z_{d}(n)$ due to delay spread of $h_{cd}(n)$ and misalignment $T$ comprises the deviation from the cyclic assumption. Note that this fold-in happens both at the start and end of the $P$ sample DFT window. Using \eqref{eqn:conv} and the decomposition $z_{c}(n)=z_{c-cyc}(n)+z_{c1}(n)+z_{c2}(n)$ we can see that: 

\begin{flalign}
z_d(n)=\sum_{k=0}^{k=L-1}\left(z_{c-cyc}(n-k)h_{cd}(k)\right)\,+\,d(n), 0 \leq  n \leq P-1
\end{flalign}

\begin{flalign} \label{eqn:uncancellable}
d(n) &=\sum_{k=T+1}^{L-1-n}{\left(z_{c}(-k)-z_{c}(P-k)\right) h_{cd}(k+n)}\\ \nonumber 
&0 \leq n \leq L-T-2  \label{eqn:uncancellable2}
\end{flalign}
\begin{flalign}
d(n) &=\sum_{k=0}^{n-(P-T)} \left(z_{c}(n-k)-z_{c}(n-k-P)\right) h_{cd}(k) \\ \nonumber 
&P-T \leq n \leq P-1 \\ \nonumber
d(n) &= 0, \text{otherwise}
\end{flalign}

Considering the DFT of the signal $z_d(n)$, we get:
\begin{flalign}{\label{eqn:total_noise}}
Z_{d}(q)=Z_c(q)\sum_{k=0}^{k=L-1}h_{cd}(k)W_{P}^{(k-T)q}+D(q)
\end{flalign}
\begin{flalign}{\label{eqn:fd_uncancellable1}}
D(q) = \sum_{n=0}^{L-T-2}{\sum_{k=T+1}^{L-1-n}{\left(z_{c}(-k)-z_{c}(P-k)\right) h_{cd}(k+n)}W_{P}^{nq}}\\ 
+\sum_{n=P-T}^{P-1}{\sum_{k=T+1}^{L-1-n}{\left(z_{c}(n-k)-z_{c}(n-k-P)\right) h_{cd}(k+n)}W_{P}^{nq}}
\end{flalign}
Rearranging the terms in \eqref{eqn:fd_uncancellable1}, we get the following:
\begin{flalign}{\label{eqn:fd_uncancellable2}}
D(q)=\sum_{m=T+1}^{L-1}{(z_{c}(-m)-z_{c}(P-m))F_{m}(q)W_{P}^{-mq}}\\
+\sum_{m=0}^{T-1}{(z_c(P-T+m)-z_c(-T+m))G_{m}W_{P}^{(P-T+m)q}}
\end{flalign}
where $F_{m}(q)=\sum_{i=0}^{L-1-m}h_{cd}(m+i)W_{P}^{(m+i)q}$ and $G_{m}(q)=\sum_{i=0}^{m}h_{cd}(i)W_{P}^{iq}$ are DFT's of rectangular
windowed versions of $h(n)$. Further rearranging \eqref{eqn:fd_uncancellable2}, we get:
\begin{flalign}\label{eqn:fd_uncancellable3}
D(q)&= \sum_{m=T+1}^{L-1}{R_{m}(q)h_{cd}(m)W_{P}^{mq}}+\sum_{m=0}^{T-1}{S_{m}(q)h_{cd}(m)W_{P}^{mq}} \\ \label{eqn:fd_uncancellable4}
R_{m}(q)&=\sum_{k=m}^{L-1}\left(z_c(-k)-z_c(P-k)\right)W_{P}^{(P-k)q} \\ \label{eqn:fd_uncancellable5}
S_{m}(q)&=\sum_{k=0}^{m}\left(z_c(P-T+k)-z_c(-T+k)\right)W_{P}^{(-T+k)q} 
\end{flalign}
$R_m(q)$ and $S_m(q)$ are DFTs of rectangular windowed versions of the alien noise signal and can alternately be written as a convolution the DFTs of $z_c(n)$ and the rectangular window. This leads to the following:
\begin{flalign} \label{eqn:fd_uncancellable6}
R_{m}(q)&=\frac{1}{P}\sum_{k=0}^{P-1}Z_{c}^{start-res}(q-k)B_{m}(k)
\end{flalign}
\begin{flalign} \label{eqn:fd_uncancellable7}
S_{m}(q)&=\frac{1}{P}\sum_{k=0}^{P-1}Z_{c}^{end-res}(q-k)C_{m}(k)
\end{flalign}
where: 
\begin{flalign} \label{eqn:fd_uncancellable8}
Z_c^{start-res}(q)=\sum_{k=-T}^{k=P-T-1}(z_c(k-P)-z_c(k))W_{P}^{kq}
\end{flalign}
\begin{flalign} \label{eqn:fd_uncancellable9}
Z_c^{end-res}(q)=\sum_{k=-T}^{k=P-T-1}(z_c(k+P)-z_c(k))W_{P}^{kq}
\end{flalign}
and $B_{m}(k)$ and $C_{m}(k)$ correspond to DFTs of rectangular windows and are given by:
\begin{flalign}
B_{m}(q)\,&=\,\sum_{k=P-m}^{k=P-T-1}W_{P}^{kq},\,\, T+1 \leq m \leq L-1 \\
C_{m}(q)\,&=\,\sum_{k=-T}^{-m-1}W_{P}^{kq},\,\, 0 \leq m \leq T-1
\end{flalign}

The coefficient $H_{pertone}(q)$ is given by $H_{pertone}(q)=\frac{E\{Z_d(q)Z_c^{*}(q)\}}{E\{Z_c(q)Z_c^{*}(q)\}}$. Using  \eqref{eqn:total_noise}, we get $H_{pertone}(q)=\sum_{k=0}^{L-1}h_{cd}(k)W_{P}^{(k-T)q}+\frac{E\{D(q)Z_c^{*}(q)\}}{E\{Z_c(q)Z_c^{*}(q)\}}$. Using \eqref{eqn:fd_uncancellable3}, we get:
\begin{flalign} \label{eqn:result_pre}
H_{pertone}(q)&=\sum_{k=0}^{L-1}h_{cd}(k)W_{P}^{(k-T)q}\,+\\ \nonumber
&\frac{\sum_{m=T+1}^{L-1}{E\{Z_c^{*}(q)R_{m}(q)\}h(m)W_{P}^{mq}}}{{E\{Z_c(q)Z_c^{*}(q)\}}}\,+\\\nonumber
&\frac{\sum_{m=0}^{T-1}{E\{Z_c^{*}(q)S_{m}(q)\}h(m)W_{P}^{mq}}}{{E\{Z_c(q)Z_c^{*}(q)\}}}
\end{flalign}
Assuming $z_c(n)$ is \textit{wide-sense stationary} and \textit{white}, using \eqref{eqn:fd_uncancellable6},\eqref{eqn:fd_uncancellable7}, \eqref{eqn:fd_uncancellable8},\eqref{eqn:fd_uncancellable9} and noting that $B_{m}(0)=m-T$,$C_{m}(0)=T-m$, \eqref{eqn:result_pre} reduces to:
\begin{flalign} \label{eqn:result_pertone}
H_{pertone}(q)&=\sum_{k=0}^{k=T-1}h_{cd}(k)\left(1+\frac{k-T}{P}\right)W_{P}^{(k-T)q}\\\nonumber
&+ \sum_{k=T}^{k=L-1}h_{cd}(k)\left(1-\frac{k-T}{P}\right)W_{P}^{(k-T)q} 
\end{flalign}  

From \eqref{eqn:result_pertone} it can be seen that the inverse DFT of per-tone coefficients $H_{pertone}(q)$ yields a cyclically shifted ( by $T$ samples ) time domain signal $h_{pertone}(n)$ which is related to the true impulse response $h_{cd}(n)$ cyclically shifted by $T$ samples. Therefore we have proven the following result:

\begin{lemma}\label{lemma_pertone}
If $z_c(n)$ is assumed to be white and wide-sense stationary and $T<L$ and $T \geq 0$(i.e. CM DFT window is ahead in time w.r.t DM DFT window), $h_{pertone}(n)$ derived from the inverse DFT of the per-tone coefficients $H_{pertone}(q)$ is related to the true impulse response $h_{cd}(n)$ as:
\begin{flalign}
h_{pertone}(i) &= h_{cd}^{(cyc-T)}(i) \left(1-\frac{i}{P} \right), \,  i = 0,1, \dots L-T \\
h_{pertone}(P+i) &= h_{cd}^{(cyc-T)}(P+i) \left(1+\frac{i}{P} \right), \, i = -1, \dots -T
\end{flalign}
where $h_{cd}^{cyc-T}$ s the cyclically shifted version of $h_{cd}(n)$.
\end{lemma}
Corresponding results can also be derived for the scenarios where $z_c(n)$ is coloured or $T>L$ or $T<0$ (i.e. CM DFT window lags the DM DFT window) but discussing all these scenarios is beyond the scope of this paper.
\par
This result enables us to derive the true impulse response of the coupling function from the per-tone coefficients, which can be  can be estimated quickly and with low complexity.  The true impulse response can derive an optimum time-domain canceller ( e.g. based on the MMSE criterion ) which does not suffer from the loss in per-tone cancellation or can be used for a post-training "delay" adjustment to optimize the training complexity.

\subsection{Optimizing per-tone cancellation via delay adjustment}

The uncancellable residual time domain signal with per-tone cancellation is given by:
\begin{flalign}
r(n)=d(n)+\sum_{k=0}^{L-1}\left(h_{cd}(k)-h_{pertone}(k)\right)z_{c-cyc}(n-k)
\end{flalign}
The total uncancellable noise energy in a DMT symbol given by $\xi(\boldsymbol{h},T)=\sum_{q=0}^{q=P-1}I_{U}(q)=\sum_{n=0}^{P-1}{E[r(n)^{2}]}$.
\begin{flalign} \label{eqn:residual}
\xi(\boldsymbol{h},T) = 2\left( \sum_{m=1}^{L-T-1} \left( \boldsymbol{h}_{m}^{'T} \mathbf{R}_{m} \boldsymbol{h}_{m}^{'} - \boldsymbol{h}_{m}^{'T} \mathbf{P}_{m} \boldsymbol{h}_{m}^{'} \right) + 
\sum_{t=0}^{T-1} \left(\boldsymbol{h}_{t}^{''T} \mathbf{R}_{t} \boldsymbol{h}_{t}^{''} - \boldsymbol{h}_{t}^{''T} \mathbf{P}_{t} \boldsymbol{h}_{t}^{''} \right)  \right)     
\end{flalign}

where $ \boldsymbol{h}_{m}^{'}=\begin{bmatrix}
h(T+m)\\
..\\
h(L-1) 
\end{bmatrix}$ and $ \boldsymbol{h}_{t}^{''} =\begin{bmatrix}
h(0)\\
..\\
h(t) 
\end{bmatrix}$ are rectangular windowed versions of the impulse response $h(n)$ while 
$\mathbf{R}_{m}$ and $\mathbf{R}_{t}$ 
are $(L-T-m)\times(L-T-m)$ and $(t+1)\times(t+1)$ autocorrelation matrices of $z_{c}(n)$ which is assumed to be wide-sense stationary and $\mathbf{P_{t}} = \begin{bmatrix}
r(N) & r(N+1) &  ..& r(N+t)\\
r(N-1) & r(N) &  ..& ..\\
.. & .. & .. &  ..\\
.. & .. &  ..& r(N)
\end{bmatrix}$ and $\mathbf{P_{m}} = \begin{bmatrix}
r(N) & r(N+1) &  ..& r(N-L+m+T+1)\\
r(N-1) & r(N) &  ..& ..\\
.. & .. & .. &  ..\\
.. & .. &  ..& r(N)
\end{bmatrix}$ matrices also consist of the autocorrelation terms of $z_{cm}(n)$.

The above result assumes that
$P>>L$ and the autocorrelation terms $r(n)=E\{z_{c}(i)z_{c}(i+n)\} \approx 0$ for $n \approx P$ since $P$ (length of DMT symbol) is very large.   This method of estimating the uncancellable noise is similar to the ISI-ICI analysis presented in \cite{henkel2002cyclic}. 

\par The uncancellable noise energy depends on the spectral characteristics of the alien noise signal $z_{c}(n)$, the delay spread of $h(n)$ as well as the misalignment $T$ between the CM and DM DFT windows. As long as the length of DMT symbol $P$ (e.g. $P=8192$ samples for VDSL) is significantly longer than the delay spread of $h(n)$, the span of $u(n)$ remains small in comparison to $P$ and consequently the uncancellable noise energy remains small. The contribution to the uncancellable energy due the misalignment $T$ can be minimized by optimizing w.r.t $T$ as shown next. 
\par
\textit{Remark:} We cannot minimize the uncancellable noise power $I_U(q)$ on each tone since we have only one adjustable parameter i.e. $T$ and instead attempt to minimize the sum of $I_U(q)$ over all the tones ($\sum_{q=0}^{P-1}I_U(q)$) by adjusting $T$.

\subsection{Optimization of Uncancellable Energy w.r.t. Misalignment}
The total uncancellable energy per DMT symbol can be minimized by considering the following optimization problem in terms of $T$:  
\begin{equation} \label{eqn:opt_T}
T_{opt}=\underset{T}{\argmin}{\,\xi(\boldsymbol{h},T) }
\end{equation}
where $\xi(\boldsymbol{h},T)$ is given by \eqref{eqn:residual}. This optimization corresponds to selecting a misalignment so as to minimize the energy in the tails of $h(n)$, which in turn will reduce the signal folding in into the DM DFT window from outside the CM DFT window. Without this optimization, the uncancellable energy in the per-tone model may be significantly higher. To perform this optimization, we need to estimate the true coupling function $h(n)$.
\par
The per-tone canceller coefficient estimation happens with an arbitrary misalignment $T_{trg}$ using the method(s) described in Section \ref{sec:anc}. The misalignment $T_{trg}$ during canceller training as well as the length $L$ (containing a significant portion of the energy in the impulse response) can be identified from the inverse DFT $h_{cd}(n)$ of the estimated per-tone coefficients $H_{cd}(q) = \beta(q)(1+\delta(q))$. Once the per-tone coefficients and the misalignment $T_{trg}$ are estimated, the true impulse response $h(n)$ can be estimated by using Lemma \ref{lemma_pertone} (stated without proof).

Having estimated $h(n)$, $T_{opt}$ can be evaluated by considering \eqref{eqn:opt_T}. The CM DFT window can then be shifted by the delay $(T_{opt}-T_{trg})$ and the new per-tone coefficients calculated using Lemma \ref{lemma_pertone}. The optimization procedure described  is independent of the method used for training the per-tone canceller and can be conveniently applied as a post-training adjustment resulting in a modified position for the CM DFT window and new values for the per-tone coefficients.

\subsection{Measurement Results}
\begin{figure}
\includegraphics[clip, trim=0.5cm 9.5cm 0.5cm 9.5cm, width=0.5\textwidth]{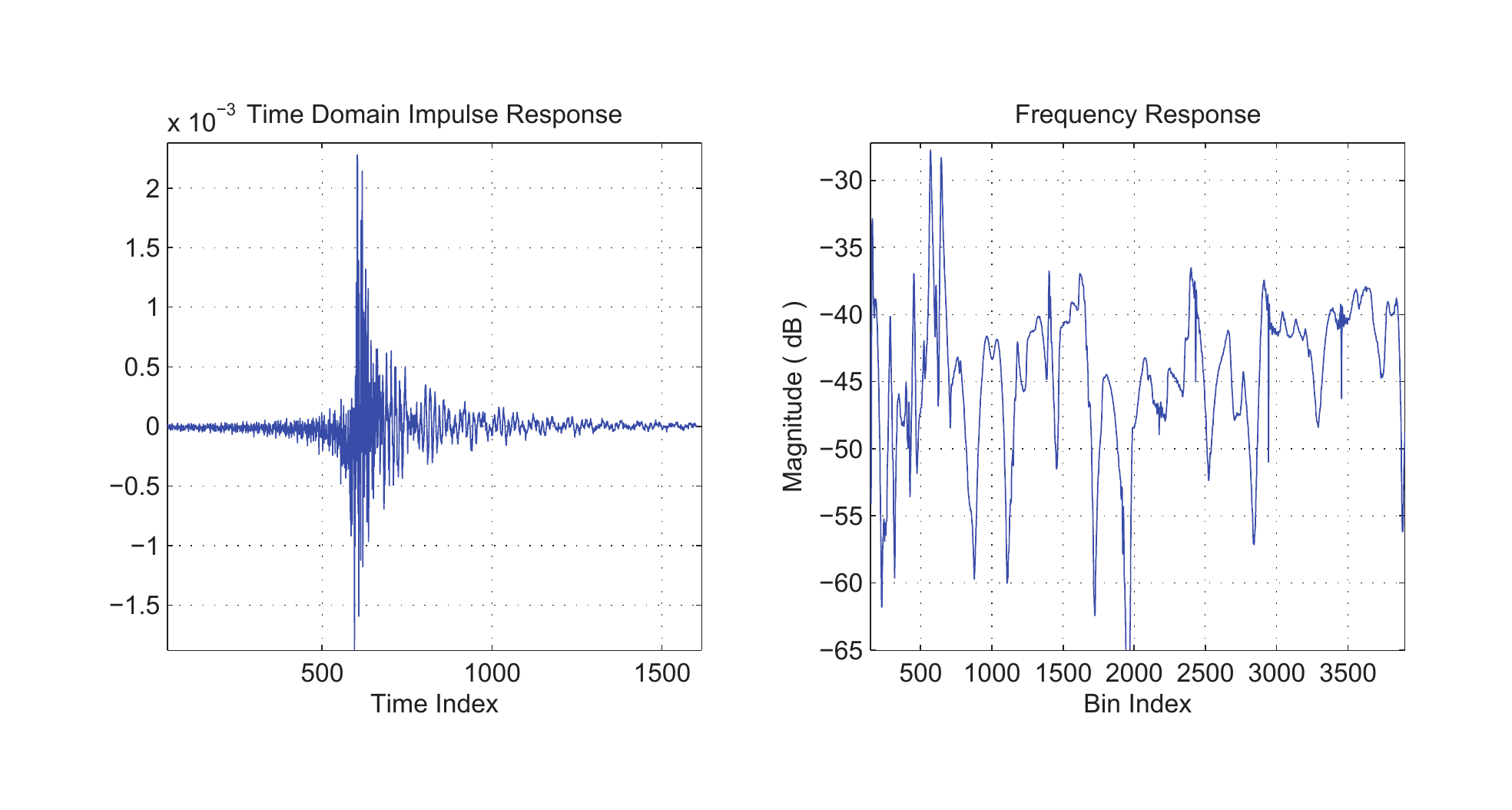}
\caption{CM-DM coupling function measured by injecting white noise via a common mode injector into a 24AWG loop }
\label{fig:coupling_function}
\end{figure}

\begin{figure}
\includegraphics[clip, trim=2.5cm 12cm 2.5cm 12cm, width=0.5\textwidth]{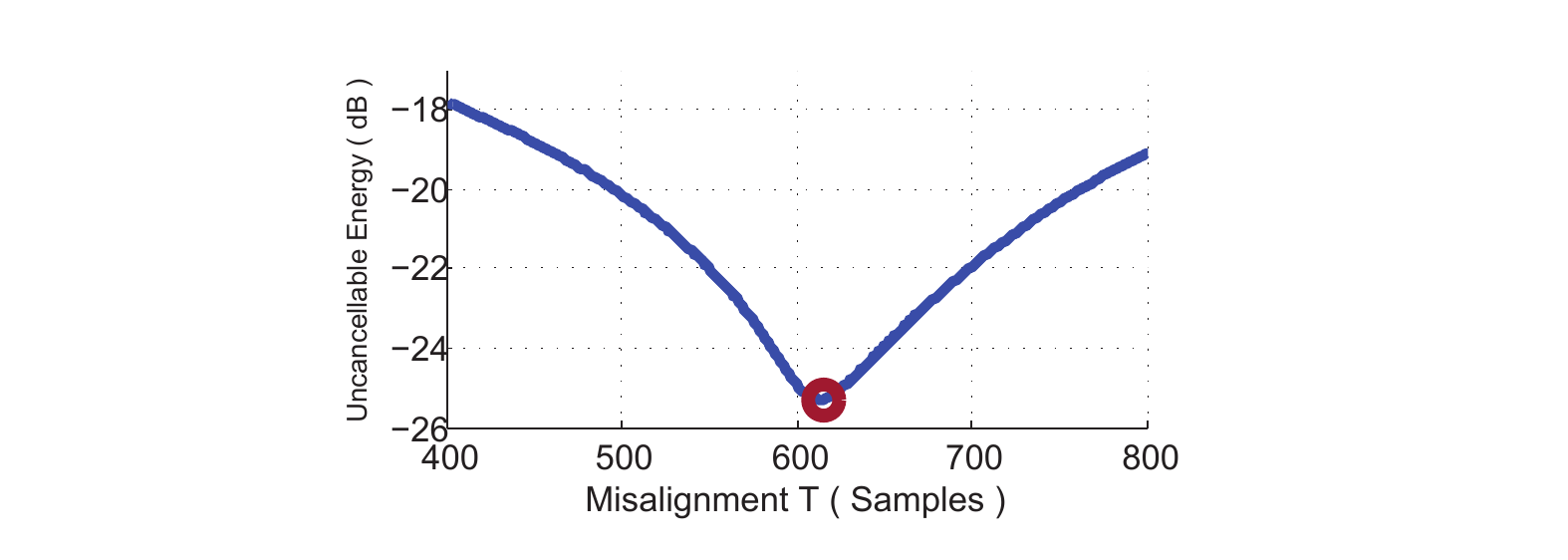}
\caption{Uncancellable Energy per DMT Symbol vs Misalignment for Impulse Response shown in Fig \ref{fig:coupling_function} }
\label{fig:uncanc_energy}
\end{figure}

\begin{figure}
\includegraphics[clip, trim=2.5cm 10cm 2.5cm 11cm, width=0.5\textwidth]{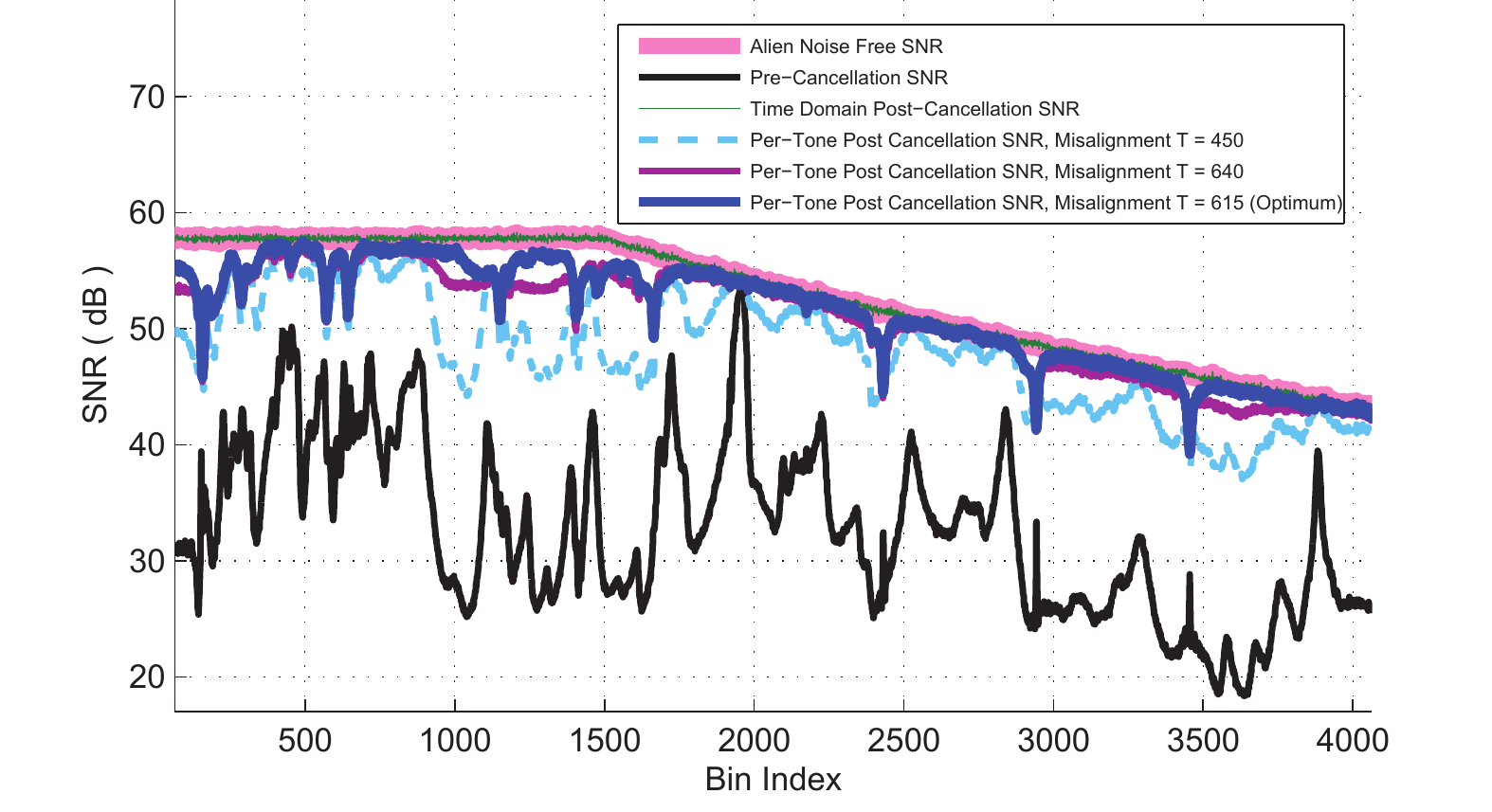}
\caption{Performance Gap between Per-Tone Cancellation and Time Domain Cancellation with alien noise modelled as stationary white noise }
\label{fig:pertone_limit_whitenoise}
\end{figure}

Figure \ref{fig:coupling_function} shows the measured CM-DM coupling function for a 400m 24 AWG loop obtained by injecting white noise using a specially designed CM injection probe and measuring the signals at the CM and DM sensors. It is seen that the delay spread of the impulse response is quite large ($\approx 700$ samples). Figure \ref{fig:uncanc_energy} shows the uncancellable energy calculate according to \eqref{eqn:residual} as a function of the misalignment $T$ (Optimum misalignment is $T_{opt}=615$). Figure \ref{fig:pertone_limit_whitenoise} shows the corresponding post-cancellation performance for per-tone (with different values of misalignment) as well as time-domain linear cancellation with alien noise modelled as stationary  white  noise(Refer table \ref{tab:simulation_setup} for the simulation parameters). It is seen that the gap between per-tone cancellation and linear cancellation is small for majority of the bins when the misalignment is optimum.
\par
\textit{Remark:} For REIN sources like home appliances the length of the impulse noise burst may be much smaller than the DMT symbol \cite{AEEE942}. For these cases, the uncancellable noise energy can be small given that start and/or end portion of the uncancellable signal $u(n)$ (refer \eqref{eqn:uncancellable},\eqref{eqn:uncancellable2}) may be zero due to the shorter length of the noise burst.


\section{Conclusion}

\bibliographystyle{IEEEbib}
\bibliography{refs}

\end{document}